\documentclass[%
 aip,
 jmp,
 amsmath,amssymb,
 reprint,%
]{revtex4-1}

\usepackage{graphicx}
\usepackage{dcolumn}
\usepackage{bm}

\usepackage[utf8]{inputenc}
\usepackage[T1]{fontenc}
\usepackage{etoolbox}
\usepackage{mathrsfs}
\usepackage{yfonts}
\usepackage{xcolor}

\makeatletter
\def\@email#1#2{%
 \endgroup
 \patchcmd{\titleblock@produce}
  {\frontmatter@RRAPformat}
  {\frontmatter@RRAPformat{\produce@RRAP{*#1\href{mailto:#2}{#2}}}\frontmatter@RRAPformat}
  {}{}
}%
\makeatother
\begin{document}

\preprint{AIP/123-QED}

\title{Ready-to-fit inhomogeneous cosmological model: The axially symmetric Szekeres spacetime} 



\author{Marie-No\"elle C\'el\'erier}
\affiliation{Laboratoire d’étude de l’Univers et des phénomènes eXtrèmes (LUX), Observatoire de Paris-PSL, UMR 8262 CNRS, Sorbonne Université, 5, Place Jules Janssen, F-92190 Meudon, France}
\email{marie-noelle.celerier@obspm.fr}

\date{\today}

\begin{abstract}
The purpose of the present work is based on two main observations: the tensions encountered by the standard $\Lambda$CDM model when confronted to precision small scale cosmological data and the finding that the matter distribution is dipolar and the expansion of the Universe is axially symmetric roughly in the direction of the CMB dipole. Therefore, we propose, as a model for the inhomogeneous local universe, an axially symmetric Szekeres solution. After describing its main properties, we are left with three metric functions to be fitted to data between the observer and the transition to homogeneity which is an intrinsic feature of Szekeres spacetimes. So as to turn a difficult functional inference problem into a classical parameter estimation problem, we propose to use Chebyshev polynomial expansions, which, as a first step, we truncate after the second order terms. We are thus left with nine constant parameters: six for the metric functions, plus the observer's location and the cosmological constant. Here are the proper ingredients needed to implement the data fitting to the model in the future.
\end{abstract}


\maketitle 

\section{Introduction} \label{intro}

It is now a common statement to stress that the current $\Lambda$CDM standard model, while yielding amazing results for describing our Universe at large scales, encounters tensions when confronted with the data of small scale precision cosmology. The most significant issue is the $H_0$ tension which has been recently hammered out by the H0DN Collaboration, who found a $7.1 \sigma$ discrepancy with the early universe plus $\Lambda$CDM result \cite{H026}. A number of possible solutions to this issue have been proposed in the literature \cite{D25}. However, since general relativity (GR) has successfully passed the numerous tests which have been designed to challenge it up to severe bounds, i. e., $10^{-4}$ to $10^{-20}$ \cite{W14,Y25,A25,G26}, it makes sense to stick to this theory. In its framework, the exact inhomogeneous solution, found in 1975 by Szekeres \cite{S75} appears well designed to reproduce our local Universe, while being able to recover naturally $\Lambda$CDM at large scales \cite{C24}.

Since the Szekeres model is rather complicated to use in its most general configuration, a simplifying assumption has often been made in the literature. The hypothesis of axial symmetry indeed allows one to reduce the number of arbitrary metric functions to be fitted to the observations. Therefore, it has been frequently used for numerical studies aiming at reproducing inhomogeneities in the Universe \cite{B10,M12,M22,H26}. However, the rationale for using this assumption was merely a simplification purpose.

Now, it appears, and it has been confirmed by a number of recent works, that the matter distribution is dipolar and that the expansion of the Universe exhibits an axial symmetry compatible with the direction of the CMB dipole \cite{B16,B18,S19a,S19b,C19,Se21,Si21,S22,T24,S24,Se25a,Se25b,R25,W25,Sa25,K26}. This gives us an additional and more physically grounded reason to study thoroughly the mathematical and physical properties of the axially symmetric Szekeres spacetimes in an improved way compared to previous studies. Therefore, we obtain a description of the Szekeres solutions using a number of auxiliary functions of the radial coordinate reduced to three.

Thus, once the axially symmetric Szekeres model is fully described and characterized, the second aim of this paper is to show how it can be parametrized so as to permit a proper practical fitting to cosmological data. The aim is to replace the auxiliary functions by a small number of constant parameters so as to drop constraining inference methods in favor of easier to manage parameter fitting. This is done through the use of Chebyshev expansions whose rapid convergence allows us to keep a small number of constant parameters to approximate satisfactorily the Szekeres functions. 

The paper is organized as follows. In Sec.\ref{sz}, the general properties of the Szekeres solutions are recalled. Section \ref{as} is devoted to a thorough study of axial symmetry applied to the Szekeres spacetimes with a special stress on issues linked to the occurrence of possible singularities. The propagation of light in such spacetimes is described in Sec.\ref{lp}. The proposed method for fitting the cosmological data by employing Chebyshev polynomial expansions is described in Sec.\ref{df}. Section \ref{co} is devoted to the conclusion.

\section{Szekeres solutions} \label{sz}

The Szekeres model is an exact irrotational solution to Einstein's field equations, exhibiting no Killing vector, hence no global symmetry, and gravitationally sourced by a zero-pressure perfect fluid, i. e., dust. For a more complete presentation of the solution the reader is referred to specialized textbooks \cite{BK10,P24} and references therein.

In synchronous and comoving projective coordinates, in geometric units, and in a parametrization convenient for our purpose \cite{H96}, the line element can be written as
\begin{equation}
\textrm{d}s^2 = - \textrm{d}t^2 + \frac{\left(\Phi_{,r} - \Phi E_{,r}/E\right)^2}{\epsilon - k} \textrm{d}r^2 + \frac{\Phi^2}{E^2}(\textrm{d}p^2 + \textrm{d}q^2), \label{s1}
\end{equation}
with 
\begin{equation}
E(r,p,q) = \frac{S}{2}\left[ \left(\frac{p-P}{S}\right)^2 + \left(\frac{q-Q}{S}\right)^2 + \epsilon \right],  \label{s2}
\end{equation}
where $\Phi$ is a function of $t$ and $r$ and where $k$, $S$, $P$, and $Q$ are functions of the $r$ coordinate alone. The parameter $\epsilon$ determines whether the $(p,q)$ 2-surfaces of constant $\{r,t\}$ are 2-spheres $(\epsilon= +1)$, 2-pseudospheres, i. e., hyperboloids $(\epsilon = -1)$, or 2-planes $(\epsilon=0)$.

The quasispherical and quasihyperbolic Szekeres (QSS and QHS) models exhibit a dipolelike distribution in their energy density, whose strength and orientation are determined by the dipole functions $S$, $P$, $Q$. The surfaces $\{t=\text{const},r=\text{const}\}$ of a given Szekeres model can be quasispherical in some regions and quasihyperbolic in some others, with quasiplanar boundaries. This is a nice feature, allowing us to use the model to represent a cosmic web with its voids, clusters, and filaments. This is also the reason why we will consider in the following the three kinds of geometry by keeping, as often as possible, the $\epsilon$ parameter as such, without giving it any one of the three possible numerical values. When this will be impossible, we will anyhow consider each of the three cases separately in turn.

Another set of coordinates can be used when one wants to visualize, e. g., the direction of geodesics or the dipole in the 2-spaces $\{t=\text{const},r=\text{const}\}$. These are the polar coordinates which can be obtained from the projective coordinates through transformations depending on the quasigeometry of the surfaces:

(i) for quasispheres:
\begin{equation}
(\frac{p - P}{S}, \frac{q - Q}{S}) = \cot \frac{\theta}{2} (\cos \phi, \sin \phi). \label{s2a}
\end{equation}

(ii) for quasihyperboloids:
\begin{equation}
(\frac{p - P}{S}, \frac{q - Q}{S}) = \coth \frac{\theta}{2} (\cos \phi, \sin \phi).
\end{equation}

(iii) for quasiplanes:
\begin{equation}
(\frac{p - P}{S}, \frac{q - Q}{S}) = \frac{2}{\theta} (\cos \phi, \sin \phi).
\end{equation}

The field equations with a cosmological constant reduce to
\begin{equation}
\Phi_{,t}^2 = \frac{2 M}{\Phi} - k + \frac{\Lambda}{3} \Phi^2, \label{s3}
\end{equation}
\begin{equation}
4 \pi \rho(t,r,p,q) = \frac{M_{,r} - 3 M E_{,r}/E}{\Phi^2 (\Phi_{,r} - \Phi E_{,r}/E)}, \label{s4}
\end{equation}
where $\rho$ is the matter energy density and $M(r)$ is an integration function whose interpretation varies with the quasigeometry of the considered region.

Equation (\ref{s3}) can be integrated as
\begin{equation}
t - t_B(r) = \int_0^{\Phi} \frac{\text{d}\tilde{\Phi}}{{\sqrt{\frac{2 M}{\tilde{\Phi}} - k + \frac{\Lambda}{3} \tilde{\Phi}^2}}}, \label{s5}
\end{equation}
where $t_B(r)$ is an arbitrary function which stands for the big bang or crunch time.

In the geometric optics approximation \cite{S61}, light travels on null geodesics. For the Szekeres model their equations read \cite{BK10}

\onecolumngrid

\begin{equation}
\frac{\text{d}^2 t}{\text{d}s^2} + \left( \frac{\Phi_{,tr} - \Phi_{,t} E_{,r}/E}{\epsilon - k}\right)\left(\Phi_{,r} - \Phi E_{,r}/E \right)\left(\frac{\text{d}r}{\text{d}s}\right)^2 + \frac{\Phi \Phi_{,t}}{E^2}\left[\left(\frac{\text{d}p}{\text{d}s}\right)^2 + \left(\frac{\text{d}q}{\text{d}s}\right)^2\right] = 0,  \label{s7}
\end{equation}
\begin{eqnarray}
\frac{\text{d}^2r}{\textrm{d}s^2} &+& 2\left(\frac{\Phi_{,tr} - \Phi_{,t} E_{,r}/E}{\Phi_{,r} - \Phi E_{,r}/E}\right)\frac{\text{d}t}{\text{d}s}\frac{\text{d}r}{\text{d}s} + \left(\frac{\Phi_{,rr} - \Phi E_{,rr}/E}{\Phi_{,r} - \Phi E_{,r}/E} - \frac{E_{,r}}{E} + \frac{k_{,r}}{2(\epsilon-k)}\right)\left(\frac{\text{d}r}{\text{d}s}\right)^2 \nonumber \\
&+& 2 \frac{\Phi}{E^2}\left(\frac{E_{,r}E_{,p} - E E_{,r p}}{\Phi_{,r} - \Phi E_{,r}/E}\right) \frac{\text{d}r}{\text{d}s}\frac{\text{d}p}{\text{d}s} + 2 \frac{\Phi}{E^2}\left(\frac{E_{,r}E_{,q} - E E_{,r q}}{\Phi_{,r} - \Phi E_{,r}/E}\right) \frac{\text{d}r}{\text{d}s}\frac{\text{d}q}{\text{d}s} \nonumber \\
&-&  \frac{\Phi}{E^2}\left(\frac{\epsilon - k}{\Phi_{,r} - \Phi E_{,r}/E}\right)\left[\left(\frac{\text{d}p}{\text{d}s}\right)^2 + \left(\frac{\text{d}q}{\text{d}s}\right)^2\right] = 0, \label{s8}
\end{eqnarray}
\begin{eqnarray}
\frac{\text{d}^2p}{\textrm{d}s^2} &+& 2 \frac{\Phi_{,t}}{\Phi}\frac{\text{d}t}{\text{d}s}\frac{\text{d}p}{\text{d}s} - \frac{\Phi_{,r} - \Phi E_{,r}/E}{\Phi(\epsilon - k)}\left(E_{,r}E_{,p} - E E_{,r p}\right)\left(\frac{\text{d}r}{\text{d}s}\right)^2 \nonumber \\
&+& 2 \frac{\Phi_{,r} - \Phi E_{,r}/E}{\Phi}\frac{\text{d}r}{\text{d}s}\frac{\text{d}p}{\text{d}s} - 2 \frac{E_{,q}}{E}\frac{\text{d}p}{\text{d}s}\frac{\text{d}q}{\text{d}s} + \frac{E_{,p}}{E}\left[- \left(\frac{\text{d}p}{\text{d}s}\right)^2 + \left(\frac{\text{d}q}{\text{d}s}\right)^2\right] = 0, \label{s9}
\end{eqnarray}
\begin{eqnarray}
\frac{\text{d}^2q}{\textrm{d}s^2} &+& 2 \frac{\Phi_{,t}}{\Phi}\frac{\text{d}t}{\text{d}s}\frac{\text{d}q}{\text{d}s} - \frac{\Phi_{,r} - \Phi E_{,r}/E}{\Phi(\epsilon - k)}\left(E_{,r}E_{,q} - E E_{,r q}\right)\left(\frac{\text{d}r}{\text{d}s}\right)^2 \nonumber \\
&+& 2 \frac{\Phi_{,r} - \Phi E_{,r}/E}{\Phi}\frac{\text{d}r}{\text{d}s}\frac{\text{d}q}{\text{d}s} - 2 \frac{E_{,p}}{E}\frac{\text{d}p}{\text{d}s}\frac{\text{d}q}{\text{d}s} + \frac{E_{,q}}{E}\left[\left(\frac{\text{d}p}{\text{d}s}\right)^2 - \left(\frac{\text{d}q}{\text{d}s}\right)^2\right] = 0. \label{s10}
\end{eqnarray}

In addition, we have the first integral of the geodesics, which reads
\begin{equation}
\left(\frac{\text{d}t}{\text{d}s}\right)^2 = \frac{\left(\Phi_{,r} - \Phi E_{,r}/E\right)^2}{\epsilon- k}\left(\frac{\text{d}r}{\text{d}s}\right)^2 + \frac{\Phi^2}{E^2}\left[\left(\frac{\text{d}p}{\text{d}s}\right)^2 + \left(\frac{\text{d}q}{\text{d}s}\right)^2\right]. \label{s11}
\end{equation}

\twocolumngrid

\section{Axial symmetry} \label{as}

A Szekeres model with a dipole lying along the same direction at every $r$ value is globally axially symmetric. It has been shown indeed that it possesses a Killing vector field which generates a single continuous rotational symmetry \cite{G17}.

\subsection{Conditions on the metric functions implied by axial symmetry}

Axial symmetry in Szekeres spacetimes proceeds from a number of different conditions imposed on the dipole functions $S, P, Q$. This has been proposed for the QSS case under the form of a theorem \cite{BK10} and extended subsequently to the most general case \cite{K11}. We derive here conditions obtained in a more straightforward manner for any value of $\epsilon$. This allows us to specify and correct the previous claims and to obtain a proper validation of the simplest condition, hence setting grounds for the usual way of writing the axial symmetry assumption.

On a hypersurface determined by $t = \text{const}$, the Szekeres web is composed of a set of sheets - quasispheres, quasihyperboloids or quasiplanes - which exhibit not only shifted quasicenters but also rotated dipoles when passing from one sheet to another. 
 
The shortest distance from a given spacetime point to a nearby sheet is along the path $\textrm{d}p = \textrm{d}q = 0$. If the point is on the symmetry axis, symmetry arguments imply that this minimal distance is directed along the axis. In any other direction, the symmetry would be broken. Hence, the shifting direction can be found by imposing $\textrm{d}p = \textrm{d}q = 0$ on a solution to the set of null geodesic equations (\ref{s7})--(\ref{s10}). However, for a general geodesic, even in an axially symmetric spacetime, this condition imposed on (\ref{s7})--(\ref{s10}) would imply $\textrm{d}^2 p/\textrm{d}s^2 \neq 0 \neq \textrm{d}^2 q/\textrm{d}s^2$, which would mean that the condition $\textrm{d}p = \textrm{d}q = 0$ cannot be preserved all along the curve. A null geodesic with $\textrm{d}p = \textrm{d}q = 0$ all along in an axially symmetric Szekeres model travels merely the axis itself (see below).

As follows from (\ref{s9}) and (\ref{s10}), if initially $\textrm{d}p/\textrm{d}s = \textrm{d}q/\textrm{d}s = 0$, the coordinates $p$ and $q$ remain constant only if, along the whole geodesic, the coefficient of $(\textrm{d}r/\textrm{d}s)^2$ vanishes (since the other terms in $\textrm{d}p/\textrm{d}s$ and $\textrm{d}q/\textrm{d}s$ vanish from the initial conditions). This happens only if, along the whole geodesic \cite{BK10}, either
\begin{equation}
\Phi_{,r} = \Phi E_{,r}/E, \label{s12}
\end{equation}
or
\begin{equation}
E E_{,rp} = E_{,r} E_{,p}, \quad E E_{,rq} = E_{,r} E_{,q}. \label{s13}
\end{equation}

Equation (\ref{s12}) implies a shell-crossing singularity \cite{H02}. Thus, we discard this case. Equation (\ref{s13}) gives the two extrema of $E_{,r}/E$. If these equations hold for the same $(p,q)$, which we call $(p_0,q_0)$, for all $r$, the model is axially symmetric. This model also satisfies (\ref{s13}) for a second couple ($p,q$), which we call $(p_1,q_1)$, for the part of the symmetry axis on the opposite side of the origin.

By inserting (\ref{s2}) in (\ref{s13}) and setting $p=p_0$, $q=q_0$, we obtain
\begin{eqnarray}
2 \epsilon (p_0 - P)S S_{,r} - 2 (p_0 - P)(q_0 - Q) Q_{,r} \nonumber \\
= \left[ (p_0 - P)^2 - (q_0 - Q)^2 - \epsilon S^2 \right] P_{,r}, \label{s14}
\end{eqnarray}
\begin{eqnarray}
2 \epsilon (q_0 - Q)S S_{,r} - 2 (p_0 - P)(q_0 - Q) P_{,r} \nonumber \\
= \left[ (q_0 - Q)^2 - (p_0 - P)^2 - \epsilon S^2 \right] Q_{,r}. \label{s15}
\end{eqnarray}

\subsubsection{Case $P_{,r} = Q_{,r} = 0$}

Equations (\ref{s14}) and (\ref{s15}) are trivially solved by $P_{,r} = Q_{,r} = 0$, with $P = p_0$ and $Q = q_0$. In this case, the symmetry axis passes through $\theta = 0$ and $\theta = \pi$ on all sheets, with $\theta = \pi$ forwarded at infinity in the QHS case.

Instead, along symmetry axes found in more general cases, the spherical coordinates, $\theta, \phi$, do not stay constant, while the projective coordinates, $p, q$, do.  This proceeds from the diagonality of the metric in projective coordinates \cite{B20}.

\subsubsection{Case $Q_{,r} = 0$ and $P_{,r} \neq 0$}

If $Q_{,r} = 0$ and $P_{,r} \neq 0$, the axial symmetry conditions (\ref{s14}) and (\ref{s15}) reduce to
\begin{equation}
2 \epsilon (p_0 - P)S S_{,r} = \left[ (p_0 - P)^2 - (q_0 - Q)^2 - \epsilon S^2 \right] P_{,r} \label{s16}
\end{equation}
and
\begin{equation}
 \epsilon (q_0 - Q)S S_{,r} - (p_0 - P)(q_0 - Q) P_{,r} = 0. \label{s17}
\end{equation}

Equation (\ref{s17}) has two solutions. One is
$\epsilon S S_{,r} = (p_0 - P)P_{,r}$. Inserting this constraint into (\ref{s16}), we obtain either $P_{,r} = 0$ and thus $S_{,r} = 0$, or
\begin{equation}
(p_0 - P)^2 + (q_0 - Q)^2 + \epsilon S^2 = 0. \label{s18}
\end{equation}
The first solution is a mere Lema\^itre-Tolman-Bondi model and (\ref{s18}) possesses real-valued solutions only if $S = 0$. This would imply singularities in the metric and in the energy density which result in an invalid solution.

We are therefore left with the $Q= q_0$ solution of (\ref{s17}) which, inserted into (\ref{s16}), gives
\begin{equation}
2 \epsilon (p_0 - P)S S_{,r} = \left[ (p_0 - P)^2- \epsilon S^2 \right] P_{,r}, \label{s19}
\end{equation}
which can be integrated by
\begin{equation}
\epsilon S^2 = (p_0 - P)(C_2 - p_0 + P), \label{s20}
\end{equation}
where $C_2$ is an arbitrary integration constant.

A similar solution is obtained for $P_{,r} = 0$ and $Q_{,r} \neq 0$, by replacing $p_0$ and $P$ with $q_0$ and $Q$.

\subsubsection{Case $P_{,r} \neq 0$ and $Q_{,r} \neq 0$}

Now, we consider the most general case where $P_{,r} \neq 0$ and $Q_{,r} \neq 0$. From (\ref{s13}) we have \cite{BK10}
\begin{equation}
\frac{E_{,rp}}{E_{,p}} = \frac{E_{,rq}}{E_{,q}}, \label{s21}
\end{equation}
where we insert the expression of $E$ and obtain
\begin{equation}
\frac{P_{,r}}{p_0 - P} = \frac{Q_{,r}}{q_0 - Q}, \label{s22}
\end{equation}
which can be integrated with respect to $r$ as
\begin{equation}
Q = C_0 P + (q_0 - C_0 p_0), \label{s23}
\end{equation}
where $C_0$ is an integration constant. By inserting (\ref{s22}) and (\ref{s23}) into (\ref{s14}), we obtain
\begin{equation}
2 \epsilon (p_0 - P)S S_{,r}  = \left[ (1 + C_0^2)(p_0 - P)^2 - \epsilon S^2 \right] P_{,r}, \label{s24}    
\end{equation}
whose solution can be written as
\begin{equation}
\epsilon S^2 + \left(1 + C_0^2 \right)P^2 = \left[2 p_0 \left(1 + C_0^2 \right) - C_1\right] P+ C_3, \label{s25}
\end{equation}
where $C_1$ and $C_3$ are other arbitrary constants.

Now, following Appendix B of \cite{K11}, we use the below properties of the general Szekeres metrics. It is easy to verify that the metric does not change form under the following coordinate transformations:

(1) translations:
\begin{equation}
(p,q) = (p' + p_0, q' + q_0), \label{s25b}
\end{equation}
where $p_0$ and $q_0$ are arbitrary constants. Therefore, $P$ and $Q$ are transformed in $\tilde{P}$ and $\tilde{Q}$ written as
\begin{equation}
(\tilde{P},\tilde{Q}) = (p_0 - P, q_0 - Q). \label{s25c}
\end{equation}

(2) orthogonal transformations:
\begin{equation}
p = \frac{ap' + bq'}{\sqrt{a^2 + b^2}}, \quad q = \frac{-bp' + aq'}{\sqrt{a^2 + b^2}}, \label{s25d}
\end{equation}
implying a change of $(P,Q)$ to
\begin{equation}
\tilde{P} = \frac{aP - bQ}{\sqrt{a^2 + b^2}}, \quad \tilde{Q} = \frac{bP + aQ}{\sqrt{a^2 + b^2}}. \label{s25e}
\end{equation}

(3) conformal symmetries of a Euclidean 2-plane, otherwise called two-dimensional Haantjes transformations \cite{P24}:
\begin{equation}
p = \frac{p' + \lambda_1 (p'^2 + q'^2)}{T},  \label{s25f}
\end{equation}
\begin{equation}
q = \frac{q' + \lambda_2 (p'^2 + q'^2)}{T},  \label{s25g}
\end{equation}
where
\begin{equation}
T = 1 + 2 \lambda_1 p' + 2 \lambda_2 q' + (\lambda_1^2 + \lambda_2^2)(p'^2 + q'^2), \label{s25h}
\end{equation}
with $\lambda_1$ and $\lambda_2$ arbitrary constants. Properties of this transformation, useful for our calculations, are
\begin{equation}
p^2 + q^2 = \frac{p'^2 + q'^2}{T},  \label{s25ia}
\end{equation}
\begin{equation}
\textrm{d}p^2 + \textrm{d}q^2 = \frac{\textrm{d}p'^2 + \textrm{d}q'^2}{T^2}.  \label{s25ib}
\end{equation}
Under this transformation, $P$, $Q$, and $S$ change, respectively, to
\begin{equation}
\tilde{P} = \frac{P - \lambda_1 (P^2 + Q^2 + \epsilon S^2)}{U},  \label{s25j}
\end{equation}
\begin{equation}
\tilde{Q} = \frac{Q - \lambda_2 (P^2 + Q^2 + \epsilon S^2)}{U},  \label{s25k}
\end{equation}
\begin{equation}
\tilde{S} = \frac{S}{U},  \label{s25l}
\end{equation}
with
\begin{equation}
U = 1 - 2 \lambda_1 P - 2 \lambda_2 Q + (\lambda_1^2 + \lambda_2^2)(P^2 + Q^2 + \epsilon S^2). \label{s25m}
\end{equation}

The properties displayed above will now be used to draw consequences of (\ref{s23}) and (\ref{s25}) for the general axially symmetric spacetimes.

The term in parentheses in (\ref{s23}) can be set to vanish by the transformation (\ref{s25b}) when we chose $(p_0, q_0) = (0, q_0 - C_0 p_0)$. The constant $C_0$ in (\ref{s23}) can be set to zero by the transformation (\ref{s25d}) with $b = -a C_0$. Hence, we obtain, after both transformations, $Q = 0$ in (\ref{s23}). Finally, the coefficient of $P$ in (\ref{s25}) is set to vanish by the transformation (\ref{s25b}) with $(p_0, q_0) = (C_1/2 - p_0, 0)$, where we have set $C_0 = 0$. Therefore, with no loss of generality, we can assume
\begin{equation}
q_0 =  C_0 = C_1/2 -p_0 = 0. \label{constr}
\end{equation}

We consider now the following combination of the transformation (\ref{s25b}) with the transformation (\ref{s25f}) and (\ref{s25g}):
\begin{equation}
p = p_0 + \frac{p' + \lambda_1 (p'^2 + q'^2)}{T}, \quad q = q', \label{s25n}
\end{equation}
which gives the generalization of (\ref{s25j})--(\ref{s25m}), with $\lambda_2$, so far arbitrary, here vanishing, as
\begin{equation}
\tilde{P} = \frac{P - p_0 -\lambda_1 \left[(P - p_0)^2 + Q^2 + \epsilon S^2\right]}{\cal{U}},  \label{s25o}
\end{equation}
\begin{equation}
(\tilde{Q}, \tilde{S}) = (Q, S)/\cal{U},  \label{s25p}
\end{equation}
with
\begin{equation}
{\cal{U}} = 1 - 2 \lambda_1 (P - p_0) + \lambda_1^2\left[(P - p_0)^2 + Q^2 + \epsilon S^2\right]. \label{s25q}
\end{equation}
Then, using (\ref{s23}) and (\ref{s25}), with (\ref{constr}) implemented, in (\ref{s25o})--(\ref{s25q}), these become
\begin{equation}
\tilde{P} = \frac{P - p_0 -\lambda_1 (- 2 p_0 P + p_0^2 + C_3)}{\cal{U}}, \qquad \tilde{Q} = 0\label{s25r}.
\end{equation}
Now, if the constants $(p_0, \lambda_1, C_3)$, so far arbitrary, satisfy
\begin{equation}
1 + 2 \lambda _1 p_0 = 0, \quad C_3 = p_0^2,  \label{s25s}
\end{equation}
then $\tilde{P} = \tilde{Q} = 0$, and, in the $(p',q')$ coordinates, the Szekeres metric is axially symmetric. Contrary to what is advocated in \cite{K11}, the set (\ref{s25s}) has a solution whatever the value of $p_0$.

We have therefore demonstrated that the case $P_{,r} \neq 0, Q_{,r} \neq 0$ describes an axially symmetric Szekeres spacetime, whatever the value of $\epsilon = \pm 1, 0$, provided one can find a coordinate transformation yielding $\tilde{P} = \tilde{Q} = 0$ and preserving the form of the general Szekeres metric. Since GR is a covariant theory, we can, without loss of generality, state that axial symmetry in the Szekeres solution is readily defined by $P = Q = 0$ and that this constraint imposes the coordinate system in which axial symmetry can be analyzed the most easily. The four cases considered in Theorem 3.1 of \cite{BK10} can all reduce to $P = Q = 0$ provided the proper coordinate frame is used.

Returning to a general axially symmetric Szekeres model with $P_{,r} \neq 0 \neq Q_{,r}$, we recall that the dipole functions must satisfy (\ref{s23}) and (\ref{s25}) for all $r$, i. e., in the whole spacetime. These equations do not apply in the case $P = 0 = Q$. In this case, there is no constraint on the function $S$.

\subsection{Singularities} 

\subsubsection{QSS case}

In QSS models, the coordinate system defined by the symmetry axis passing through the spherical coordinate values $\theta = 0$ and $\theta = \pi$ on all shells corresponds to the case $P_{,r} = 0 = Q_{,r}$. It has been suggested by Buckley and Schlegel \cite{B20} that, in this case, a possible issue might occur, i. e., that the $p$ and $q$ coordinates diverge on half of this symmetry axis. 

The aim of this subsection is to show that such a divergence is a mere coordinate singularity and can thus be ignored for any physical purpose.

In a QSS model, antipodal points are present on any $(p,q)$ 2-surface. In spherical coordinates, the antipodal points 1 and 2 are defined by \cite{S15}
\begin{equation}
\theta_2 = \pi - \theta_1, \quad \phi_2 = \pi + \phi_1. \label{sa}
\end{equation}
The corresponding projective coordinates are given by
\begin{equation}
p_1 - P =  S \cot \frac{\theta_1}{2}\cos \phi_1, \quad q_1 - Q =  S \cot \frac{\theta_1}{2}\sin \phi_1, \label{sb}
\end{equation}
\begin{equation}
p_2 - P =  S \cot \frac{\theta_2}{2}\cos \phi_2, \quad q_2 - Q =  S \cot \frac{\theta_2}{2}\sin \phi_2, \label{sc}
\end{equation}
which can be written as
\begin{eqnarray}
p_2 - P = -S \frac{\sin \theta_1}{1 + \cos \theta_1}\cos \phi_1, \\ \nonumber
q_2 - Q = -S \frac{\sin \theta_1}{1 + \cos \theta_1}\sin \phi_1.
\label{sf}
\end{eqnarray}
Some manipulations give
\begin{equation}
\frac{\sin \theta_1}{1 + \cos \theta_1} = \pm \frac{S}{\sqrt{(p_1 - P)^2 + (q_1 - Q)^2}}. \label{sg}
\end{equation}
A straightforward calculation using the above equations yields 
\begin{equation}
\cos \phi_1 = \pm \frac{p_1 - P}{\sqrt{(p_1 - P)^2 + (q_1 - Q)^2}}. \label{sh}
\end{equation}
By inserting (\ref{sg}) and (\ref{sh}) in (51), we obtain \cite{B20}
\begin{eqnarray}
p_2 - P = -\frac{S^2 (p_1 - P)}{(p_1 - P)^2 + (q_1 - Q)^2}, \\ \label{si}
q_2 - Q = -\frac{S^2 (q_1 - Q)}{(p_1 - P)^2 + (q_1 - Q)^2},
\nonumber
\end{eqnarray}
which are the equations giving the expressions of the $(p_2,q_2)$ projective coordinates of the antipodal points with respect to $(p_1, q_1)$.

Now, we assume the axial symmetry condition $P_{,r} = 0 = Q_{,r}$ to which we add the junction to Friedmann-Lema\^itre-Robertson-Walker (FLRW) condition $P = 0 = Q$ \cite{C24}, which we insert in (54) and obtain
\begin{eqnarray}
p_2 = -\frac{p_1 S^2}{p_1^2 + q_1^2}, \\ \nonumber
q_2 = -\frac{q_1 S^2}{p_1^2 + q_1^2},
\label{sj}
\end{eqnarray}
which imply that $p_2$ and $q_2$ diverge when $p_1 = 0 = q_1$. In this case we have from (\ref{s2a})
\begin{eqnarray}
0 = \cot \frac{\theta_1}{2} \cos\phi_1, \\ \nonumber
0 = \cot \frac{\theta_1}{2} \sin\phi_1,
\label{sk}
\end{eqnarray}
which yields, for $\theta_1 = \pi$, the trivial equality $0 = 0$, whatever the value of $\phi_1$, but for $\theta_1 = 0$, an impossible equality since $\cot (0)$ diverges. Now, this does not imply that half of the symmetry axis is singular, but merely that the projective coordinates are ill-defined at the points $\{p_1, q_1\}$ corresponding to $\theta_1 = 0$, and, therefore, that the antipodal points $\{p_2, q_2\}$ are also ill-defined, but not singular, since they do not face any problem when spherical coordinates are used. Thus, as stressed by Buckley and Schlegel \cite{B20}, care must be taken when dealing with projective coordinates.

\subsubsection{Quasihyperbolic Szekeres case}

QHS models exhibit only one, not two, sheets at variance with hyperbolically symmetric spacetimes. Instead of two disjoint sheets, they display two different coordinate coverings of the same surface \cite{K12} . 

Moreover, they do not exhibit any origin where the areal radius should vanish. Therefore, no antipodal points can be defined and the corresponding issue disappears.

\subsubsection{Quasiplane Szekeres case}

It is obvious that no antipodal point can be defined in a QPS spacetime, and that the points at infinity on each planar sheet are deprived of any proper cosmological interpretation, since they are outside the Szekeres region of interest here, $r < r_{\text{trans}}$.

\section{Light propagation within axially symmetric spacetime} \label{lp}

Now that we have shown that the half-axis divergence is a mere coordinate singularity in the projective coordinates, we can use these projective coordinates while keeping this feature in mind.

\hfill

\subsection{Along the symmetry axis}

The null geodesics traveling the symmetry axis satisfy $\textrm{d}p = 0 = \textrm{d}q$, $P = 0 = Q$ and (\ref{s13}) all along. The null condition (\ref{s11}) becomes thus
\begin{equation}
\frac{\textrm{d}t}{\textrm{d}r} = \pm \frac{\Phi_{,r} - \Phi E_{,r}/E}{\sqrt{\epsilon- k}}, \label{s29}
\end{equation}
with
\begin{equation}
\frac{E_{,r}}{E} = \frac{S_{,r}}{S} \frac{\epsilon S^2 - (p^2 + q^2)}{\epsilon S^2 + (p^2 + q^2)}. \label{s30}
\end{equation}
The plus sign in (\ref{s29}) corresponds to $r_e < r_0$ and the minus sign to $r_0 < r_e$, where $r_e$ denotes the $r$ coordinate at the source and $r_0$, the $r$ coordinate at the observer.

The redshift expression is then given by \cite{BK10}
\begin{equation}
\ln (1 + z) = \pm \int^{r_0}_{r_e} \frac{\Phi_{,tr} - \Phi_{,t} E_{,r}/E}{\sqrt{\epsilon- k}} \textrm{d}r. \label{s31}
\end{equation}

\subsection{General light trajectory}

However, in an axially symmetric universe, an observer sees light rays coming from every direction, not only those following the symmetry axis. Therefore, more general equations are needed to perform a cosmological data analysis in an axially symmetric Szekeres framework.

As we have shown in Sec. \ref{as}, with no loss of generality, we can choose to work in a coordinate system where $P = 0 = Q$. Thus, the null geodesic equations (\ref{s7})--(\ref{s10}) become, with (\ref{s13}) inserted,

\onecolumngrid

\begin{equation}
\frac{\text{d}^2 t}{\text{d}s^2} + \left( \frac{\Phi_{,tr} - \Phi_{,t} E_{,r}/E}{\epsilon - k}\right)\left(\Phi_{,r} - \Phi E_{,r}/E \right)\left(\frac{\text{d}r}{\text{d}s}\right)^2 + \frac{\Phi \Phi_{,t}}{E^2}\left[\left(\frac{\text{d}p}{\text{d}s}\right)^2 + \left(\frac{\text{d}q}{\text{d}s}\right)^2\right] = 0,  \label{s32}
\end{equation}
\begin{eqnarray}
\frac{\text{d}^2r}{\textrm{d}s^2} &+& 2\left(\frac{\Phi_{,tr} - \Phi_{,t} E_{,r}/E}{\Phi_{,r} - \Phi E_{,r}/E}\right)\frac{\text{d}t}{\text{d}s}\frac{\text{d}r}{\text{d}s} + \left(\frac{\Phi_{,rr} - \Phi E_{,rr}/E}{\Phi_{,r} - \Phi E_{,r}/E} - \frac{E_{,r}}{E} + \frac{k_{,r}}{2(\epsilon-k)}\right)\left(\frac{\text{d}r}{\text{d}s}\right)^2 \nonumber \\
&-&  \frac{\Phi}{E^2}\left(\frac{\epsilon - k}{\Phi_{,r} - \Phi E_{,r}/E}\right)\left[\left(\frac{\text{d}p}{\text{d}s}\right)^2 + \left(\frac{\text{d}q}{\text{d}s}\right)^2\right] = 0, \label{s33}
\end{eqnarray}
\begin{equation}
\frac{\text{d}^2p}{\textrm{d}s^2} + 2 \frac{\Phi_{,t}}{\Phi}\frac{\text{d}t}{\text{d}s}\frac{\text{d}p}{\text{d}s}
+ 2 \frac{\Phi_{,r} - \Phi E_{,r}/E}{\Phi}\frac{\text{d}r}{\text{d}s}\frac{\text{d}p}{\text{d}s} - 2 \frac{E_{,q}}{E}\frac{\text{d}p}{\text{d}s}\frac{\text{d}q}{\text{d}s} + \frac{E_{,p}}{E}\left[- \left(\frac{\text{d}p}{\text{d}s}\right)^2 + \left(\frac{\text{d}q}{\text{d}s}\right)^2\right] = 0, \label{s34}
\end{equation}
\begin{equation}
\frac{\text{d}^2q}{\textrm{d}s^2} + 2 \frac{\Phi_{,t}}{\Phi}\frac{\text{d}t}{\text{d}s}\frac{\text{d}q}{\text{d}s} 
+ 2 \frac{\Phi_{,r} - \Phi E_{,r}/E}{\Phi}\frac{\text{d}r}{\text{d}s}\frac{\text{d}q}{\text{d}s} - 2 \frac{E_{,p}}{E}\frac{\text{d}p}{\text{d}s}\frac{\text{d}q}{\text{d}s} + \frac{E_{,q}}{E}\left[\left(\frac{\text{d}p}{\text{d}s}\right)^2 - \left(\frac{\text{d}q}{\text{d}s}\right)^2\right] = 0, \label{s35}
\end{equation}

\twocolumngrid

where $E_r/E$ is given by (\ref{s30}) and
\begin{equation}
E = \frac{p^2 + q^2}{2S} + \epsilon \frac{S}{2}, \label{s36}
\end{equation}
\begin{equation}
\frac{E_{,rr}}{E} = \frac{\epsilon S^3 S_{,rr} + (p^2 + q^2)(2 S_{,r}^2 - S S_{,rr})}{S^2\left[\epsilon S^2  + (p^2 + q^2) \right]}. \label{s37}
\end{equation}

\section{Data fitting} \label{df}

The different cosmological datasets which should be reproduced ultimately with a Szekeres model have been studied in \cite{C24}. However, for the time being, since the available data are mostly processed assuming a $\Lambda$CDM universe model, we have to reconstruct our model with the most model-agnostic strategy. To exemplify how to deal with the fitting problem in a Szekeres framework, we consider first how to manage galaxy surveys.

\subsection{Model independence}

The model independence of any field reconstruction from galaxy survey data is a strongly involved problem. Indeed, the raw data provided by full-sky and wide-area cosmological surveys are mainly the galaxy positions -- right ascension (RA), declination (Dec), and observed redshift -- galaxy shapes, spectra, angular 2-point functions, apparent magnitudes, and line-of-sight velocity dispersions. The other cosmological quantities extracted from these raw observations, such as distances, absolute magnitudes, three-dimensional (3D) clustering, and baryon acoustic oscillations (BAO) scales, are fully model dependent.

However, there are two different kinds of dependence: astrophysical and cosmological model dependence. The first one is involved, e. g., when establishing the ``raw'' distance evaluations as done in Cosmicflows-3 and --4 \cite{BT16,BT23}. Indeed, these reconstructions are done using purely astrophysical empirical relations (Tully-Fisher, Fundamental Plane, surface brightness fluctuations, Type Ia supernova light curves, etc.). These steps are mostly astrophysical calibrations and not cosmological background assumptions. Therefore, we propose to keep the results of these analyses for the fitting of the here proposed inhomogeneous cosmological model.

At variance with the unavoidable astrophysical steps, other standard assumptions made for data reconstruction will have to be avoided. As an example, as soon as one computes the peculiar velocity as
\begin{equation}
v_{pec} = c z - H_0 d, \label{vpec}
\end{equation}
one has implicitly assumed a global expansion law, a value for $H_0$, and a homogeneous background. The peculiar velocity thus obtained is cosmological model dependent. 

The Wiener filter reconstruction and its modern extensions (realizations) such as Bayesian hierarchical field-level inferences are even more constrained by the underlying cosmology, since, e. g., these inference methods rely on an assumed homogeneous cosmological model to forward the initial density field and gravity is modeled using perturbation theory or N-body simulations.

\subsection{Model degeneracy}

The axially symmetric Szekeres metric is determined by three among the four functions: $M(r)$, $k(r)$, $t_B(r)$, and $S(r)$.
One can be chosen such as to fix the $r$ coordinate freedom. We set $t_B(r) = \text{const}$, which, by a rescaling of the time coordinate, can become $t_B(r) = 0$. This gives a proper matching to FLRW at $r_{\text{trans}}$, equivalently $z_{\text{trans}}$. Moreover, it is physically motivated by large scale considerations. Indeed, inflation strongly suggests a nearly uniform bang time and CMB constraints disfavor large variations of this quantity. 

The observed radial velocities provide one scalar constraint per observed direction and redshift, while we are trying now to determine three free radial functions. The problem thus stated is underdetermined. What we observe as the purest possible is $(z, \theta, \phi)$ along our past null cone, i. e., the observed galaxy number density in redshift-angle space.

Now, in any non-FLRW spacetime (in particular, in a Szekeres spacetime), there is no unique mapping $(z, \theta, \phi) \rightarrow$ a spatial point on the $t = t_o$ surface, without solving the null geodesics of the underlying metric. Therefore, we cannot , e. g., reconstruct a present-time density field  and compare it to its Szekeres expression (\ref{s4}) without specifying in a more constrained way the spacetime geometry first.

\subsection{Szekeres fitting}

Therefore, what is the proper way for extracting an axially symmetric Szekeres model from such cosmological data? Since there is no way to infer a 3D matter density field from radial peculiar velocities in a GR framework without assuming a background expansion model, the statistical structure of the fluctuations, and a boundary behavior, a consistent strategy is to parametrize the three remaining free functions $M, k, S$, solve the null geodesics and the redshift equation, and then fit the parameters to the data. This turns a difficult functional inference problem into a classical parameter estimation problem.

The underlying undetermination can be closed here by imposing the axial symmetry (in accordance with the previously discussed observational results), the asymptotic matching to FLRW, including the choice of a vanishing bang-time function, and the parametrization of the three remaining $M(r)$, $k(r)$, and $S(r)$ functions.

Then, the strategy would be as follows: after parametrizing the three free functions, solve numerically the geodesics, compute the predicted redshift-distance relation, then the radial peculiar velocities, and fit the parameters to the survey data. This will need the writing of an appropriate numerical code, which is beyond the scope of the present work, but will have to be done later on.

Now, we wish to solve the following spectral approximation problem: approximate an unknown function on a compact interval with boundary constraints. Chebyshev polynomials of the first kind, $T_n(x)$, are known to be optimal for approximating smooth functions on finite intervals. Indeed, Chebyshev expansions minimize the maximum approximation error. They are numerically stable. They do not yield spurious singularities (unlike rational functions such as, e. g., Padé approximants) and provide an efficient truncation since lower order terms already capture the deviations, while they reduce the parameter correlations in the Markov chain Monte Carlo fits.

Thus, we will factor out the FLRW matching and the central regularity constraints before introducing free parameters through the Chebyshev polynomials.

At and beyond $r = r_{\text{trans}}$, the FLRW limit of the axially symmetric Szekeres model implies
\begin{equation}
\Phi(t,r) = r a(t), \quad M = M_0 r^3, \quad k = k_0 r^2, \quad S = 2 \epsilon. \label{trans}
\end{equation}
It includes the central regularity constraints, which impose, for $r \rightarrow 0$ \cite{BK10},
\begin{equation}
M \sim r^3, \quad k \sim r^2, \quad S \sim r^n, \quad n \geq 0, \label{reg}
\end{equation}
provided that $n = 0$.

We can therefore write each free function, collectively denoted as $F(r)$, as
\begin{equation}
F(r) = F_{\text{FLRW}} (r) D(x), \label{f1}
\end{equation}
with
\begin{equation}
x = \frac{r}{r_{\text{trans}}}. \label{f2}
\end{equation}
Then, we map the coordinate $r$ to the Chebyshev domain, through $(2x -1) \in \left[-1, +1 \right]$ and enforce the FLRW transition by imposing
\begin{equation}
D(r_{\text{trans}}) = 1. \label{f3}
\end{equation}
Since $T_n(1) = 1 \quad \forall n$, we can enforce (\ref{f3}) by writing
\begin{equation}
D(x) = 1 + \sum_{n = 1}^N a_n \left[T_n\left(2x-1\right) -1\right] , \label{f4}
\end{equation}
which implies $D(1) = 1$ as required. As $F_{\text{FLRW}} \sim F_0$ for each of the three functions, a regular center is ensured. 

Then, mapping  and applying this parametrization to each of the three free Szekeres functions, we obtain
\begin{equation}
M(r) = M_0 r^3 \left[1 + \sum_{n = 1}^N m_n \left(T_n\left(2\frac{r}{r_{\text{trans}}} -1\right) -1\right) \right], \label{f5}
\end{equation}
\begin{equation}
k(r) = k_0 r^2 \left[1 +  \sum_{n = 1}^N k_n \left(T_n\left(2\frac{r}{r_{\text{trans}}} -1\right) - 1\right) \right], \label{f6}
\end{equation}
\begin{equation}
S(r) = 2 \epsilon \left[1 +\sum_{n = 1}^N s_n \left(T_n\left(2\frac{r}{r_{\text{trans}}} -1\right) -1\right) \right]. \label{f7}
\end{equation}

The $M_0$ and $k_0$ parameters are readily fixed by the matching to FLRW at the transition scale. Implementing the correspondence between analogous terms in the Friedmann equation and in (\ref{s3}), while using (\ref{trans}), we obtain
\begin{equation}
M_0 = \frac{\Omega_b H_0^2}{2}, \quad k_0 = k ^{\Lambda\text{CDM}} = H_0^2(\Omega_b - 1) + \frac{\Lambda}{3}, \label{f8}
\end{equation}
with $\Omega_b$ the baryon density parameter of the $\Lambda$CDM standard model, $H_0$ the Hubble parameter of the same model, and $k^{\Lambda \text{CDM}}$ its curvature parameter.

Now, the problem to solve is: how many constant parameters do we need to consistently define the Szekeres model such that the reconstruction of the density field might be effective from the data without being confronted to an underdetermination problem and while converging sufficiently closely to the set of three metric functions. If each $F(2x - 1)$ is analytic, i. e., infinitely differentiable, and its Taylor series converges to it, in an ellipse in the complex plane with foci at --1 and +1 and sum of semiaxes $a + b = \rho > 1$, then the Chebyshev series converge exponentially as $\cal{O} (\rho^{-\textit{n}})$. Indeed, Chebyshev polynomials are minimax optimal for polynomial approximation on $\left[-1,+1\right]$, which means that they minimize the maximum error for a given degree $n$. The rate of convergence is determined by the size of the largest ellipse in which $F(x)$ is analytic.

Of course, the Szekeres functions that we expect to find are {\it a priori} unknown, and we cannot fix the value of $N$ from the above reasoning. However, as a first step, we could assume that the number of six parameters, which is that of the $\Lambda$CDM model, should be a good choice to begin with. It would imply $N=2$ for the degree of expansion applied to each function. Thus, we propose to complete the first numerical fittings to the cosmological data by measuring the following Szekeres parameters: $m_1$, $m_2$, $k_1$, $k_2$, $s_1$, $s_2$ defined by
\begin{eqnarray}
M(r) &=& M_0 r^3 \left[1 + m_1 \left(T_1\left(2\frac{r}{r_{\text{trans}}} -1\right) -1\right) \right.\\ \nonumber
&+& \left. m_2\left(T_2\left(2\frac{r}{r_{\text{trans}}} -1\right) -1\right)\right], \label{f9}
\end{eqnarray}
\begin{eqnarray}
k(r) &=& k_0 r^2 \left[1 + k_1 \left(T_1\left(2\frac{r}{r_{\text{trans}}} -1\right) -1\right) \right. \\ \nonumber
&+& \left. k_2\left(T_2\left(2\frac{r}{r_{\text{trans}}} -1\right) -1\right)  \right], \label{f10}
\end{eqnarray}
\begin{eqnarray}
S(r) &=& 2 \epsilon \left[1 + s_1 \left(T_1\left(2\frac{r}{r_{\text{trans}}} -1\right) -1\right) \right. \\ \nonumber
&+& \left. s_2\left(T_2\left(2\frac{r}{r_{\text{trans}}} -1\right) -1\right)\right], \label{f11}
\end{eqnarray}
with $M_0$ and $k_0$ given by (\ref{f8}). The FLRW transition parameter will be fixed {\it a priori,} depending on the resolution of the cosmological data used. 

Given that
\begin{equation}
T_1(x) = x, \quad T_2 (x) = 2x^2 - 1, \label{f12a}
\end{equation}
we can write (77)--(79) as
\begin{eqnarray}
M(r) &=& M_0 r^3 \left[1 - 2 m_1 + 2 \frac{(m_1 - 4 m_2)}{r_{\text{trans}}} r \right. \\ \nonumber
&+& \left. \frac{8 m_2}{r_{\text{trans}}^2} r^2 + \cal{O}(\textit{r}^\textit{3}) \right], \label{f12b}
\end{eqnarray}
\begin{equation}
k(r) = k_0 r^2 \left[1 - 2 k_1 + 2 \frac{(k_1 - 4 k_2)}{r_{\text{trans}}} r + \frac{8 k_2}{r_{\text{trans}}^2} r^2 + \cal{O}(\textit{r}^\textit{3}) \right], \label{f13}
\end{equation}
\begin{equation}
S(r) = 2 \epsilon \left[1 - 2 s_1 + 2 \frac{(s_1 - 4 s_2)}{r_{\text{trans}}} r + \frac{8 s_2}{r_{\text{trans}}^2} r^2 + \cal{O}(\textit{r}^\textit{3}) \right]. \label{f14}
\end{equation}
Three other parameters will have to be determined by the fittings: the observer's time and radial coordinates $(t_0,r_0)$ and the cosmological constant $\Lambda$.

Recipes for using the different cosmological datasets, either currently available or to be measured in the future, have already been displayed in a previous paper \cite{C24}, to which we refer the reader.

\section{Conclusion} \label{co}

Based on the widely accepted statement that the $\Lambda$CDM standard model of cosmology encounters tensions when compared to small scale measurements \cite{H026} and on the successes obtained by GR as a gravitation theory, we have proposed to represent the local universe by an exact GR inhomogeneous solution, i. e., the Szekeres model \cite{C24}. Indeed, Szekeres models are properly designed to feature the small scale universe since they become asymptotically homogeneous (FLRW) at large scale.

The most general Szekeres model is devoid of symmetries. However, a number of recent works have shown that the expansion of the Universe exhibit an axial symmetry roughly in the direction of the CMB dipole. This induced us to retain for our purpose the particular realization of the Szekeres model with axial symmetry, based now on physical grounds.

After recalling the main properties of the general Szekeres solution, we have displayed a thorough study of the axially symmetric cases while giving the main equations needed for a cosmological fitting. Now, as it is well-known, any axially symmetric Szekeres model can be described by only three functions of the radial coordinate $r$. However, even while this is simpler than in the most general case, the fitting of functions with datasets remains an involved issue.

Therefore, we have proposed here to parametrize these functions using Chebyshev polynomial expansions. Since Chebyshev polynomials of the first kind, $T_n(x)$, are known to be optimal for approximating smooth functions on finite intervals, and since the Szekeres models are bounded by FLRW at large scale, we have proposed to expand the three Szekeres functions to be determined, $M(r)$, $k(r)$, and $S(r)$, in Chebyshev polynomials.

We have factored out the FLRW matching together with the central regularity constraints, and then introduced free constant parameters through the Chebyshev expansion. As a first step, we have proposed to truncate this expansion at order two, which left six constant parameters to be fitted to the data instead of three functions. This has turned a difficult functional inference problem into a classical parameter estimation problem.

Hence, we have displayed here a ready-to-use method and equations to fit a robust local cosmological model with datasets. Recipes for implementing the main data surveys, either currently available or to be completed in the future, have already been given in \cite{C24}. They need the writing of numerical codes which will be the subject of future work.

Now, if ever the assumption of an axially symmetric local Universe happens not to be realized actually, it would merely imply that the number of metric functions defining the model would be increased by two; i. e., $P$ and $Q$ would no longer vanish. Then, the number of nontruncated Chebyshev parameters would be increased by four, two for each nonzero function. Moreover, the geodesic and redshift equations would recover their more involved general forms. The fitting to the data would still involve constant parameters, but in a less simplified conformation.

\section*{Data availability}

There are no publicly available research data or software supporting this manuscript. Requests for further information or data should be sent to the authors.

\end{document}